\documentclass[reprint, amsmath, amssymb, aps, superscriptaddress, prx]{revtex4-2} 

\usepackage{amsmath}
\usepackage{amssymb}
\usepackage[cjk]{kotex}
\usepackage{graphicx}
\usepackage{xcolor}
\usepackage{hyperref}
\usepackage{cancel}
\usepackage{ulem}
\usepackage{accents}
\usepackage{footmisc}
\usepackage{cancel}

\allowdisplaybreaks

\newcommand{\vacuum}{\left | \emptyset \right >}
\renewcommand{\v}{\textbf}
\newcommand{\spin}{\boldsymbol{\sigma}}

\begin{document}
\title{Dynamical defects in a two-dimensional Wigner crystal: \\ self-doping and kinetic magnetism 
}

\author{Kyung-Su Kim}
\altaffiliation{Corresponding author. \\
\href{mailto:kyungsu@stanford.edu}{kyungsu@stanford.edu}}
\affiliation{Department of Physics, Stanford University, Stanford, CA 93405}
 \author{Ilya Esterlis}
 \affiliation{Department of Physics, University of Wisconsin-Madison, Madison, Wisconsin 53706, USA}
 \author{Chaitanya Murthy} 
\affiliation{Department of Physics, Stanford University, Stanford, CA 93405}
 \author{Steven A. Kivelson}
\affiliation{Department of Physics, Stanford University, Stanford, CA 93405}

\date{\today}

\begin{abstract}
We study the quantum dynamics of interstitials and vacancies in a two-dimensional Wigner crystal (WC) using a semi-classical instanton method that is asymptotically exact at low density, i.e., in the $r_s\to \infty$ limit.
The dynamics of these point defects mediates magnetism with much higher energy scales than the exchange energies of the pure WC.
Via exact diagonalization of the derived effective Hamiltonians in the single-defect sectors, we find the dynamical corrections to the defect energies.
The resulting expression for the interstitial (vacancy) energy extrapolates to 0 at $r_s = r_{\rm mit} \approx 70$ ($r_s \approx 30$), suggestive of a self-doping instability to a partially melted WC for some range of $r_s$ below $r_{\rm mit}$.
We thus propose a ``metallic electron crystal'' phase of the two-dimensional electron gas at intermediate densities between a low density insulating WC and a high density Fermi fluid.
\end{abstract}

\maketitle

\section{Introduction}

Despite its prime importance in the field of condensed matter physics,  some basic aspects remain unsettled concerning the physics of the two-dimensional electron gas (2DEG) at intermediate densities  where various forms of ``strongly correlated electron fluids'' can arise.  The ideal 2DEG is governed by the simple Hamiltonian
\begin{align}
    \label{eq:2DEG}
    H = \sum_{i}\frac{\v p_i^{\, 2}}{2m} 
    + \sum_{i<j} \frac{e^2}{4\pi \epsilon} \frac{1}{|\v r_i -\v r_j|} ,
\end{align}
with a single dimensionless parameter,  $r_s =a_0/a_{\rm B}$,  characterizing the ratio of the typical interaction strength to the kinetic energy.
Here, $a_0 = 1/\sqrt{\pi n}$ is the average interparticle distance, $n$ is the electron density, and $a_\mathrm{B} = 4\pi \epsilon \hbar^2 /m e^2$ is the effective Bohr radius.
The phases of the 2DEG in the weak and strong coupling limits   are well-understood: it forms a paramagnetic Fermi liquid (FL) for  small $r_s$ (weak coupling) and a  Wigner crystal (WC) for large $r_s$  (strong  coupling) \cite{Wigner1934}. 
The present study addresses the intermediate coupling regime near the quantum  metal-insulator transition (MIT). 
Landmark numerical studies suggested that the MIT occurs as a direct transition from a Fermi liquid to an insulating WC at $r_s = r_{\rm melt}^* \approx 31$ \cite{tanatar1989QMC, attaccalite2002QMC, drummond2009QMC}. 
However, recent experiments \cite{hossain2020ferromagnetism,hossain2021valley,kim2021discovery,falson2022competing} suggest that the actual transition may be more complex.

\begin{figure}[b]
    \centering
    \includegraphics[width = 0.48\textwidth]{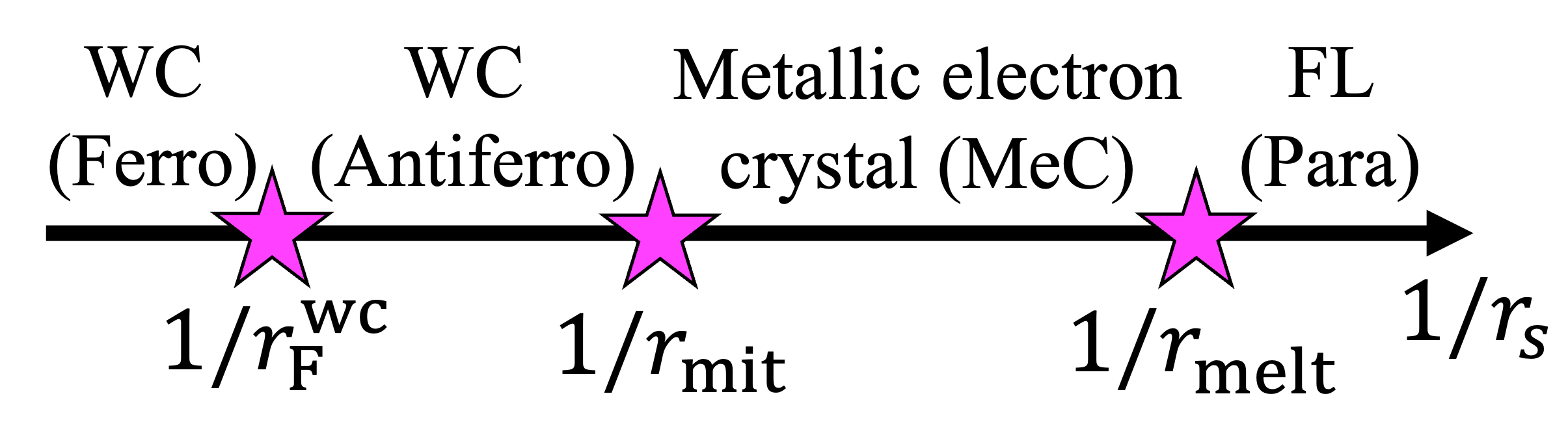}
    \caption{
    Conjectured $T=0$ phases of a clean 2DEG as a function of $1/r_s \propto \sqrt{n}$:
    WC (Ferro) = fully polarized ferromagnetic WC; 
    WC (Antiferro) = WC with some form of antiferromagnetism (or a spin liquid phase); 
    Metallic electron crystal (MeC) =  metallic electron crystal characterized by more than one electron per unit cell;
    FL (Para) = paramagnetic Fermi liquid. 
    The phase transition at $r_{\rm F}^{\rm wc} \approx 175$ \cite{Ceperley2001exchange} is due to the change of dominant exchange interactions from ferromagnetic to antiferromagnetic and is likely to be first order.
    $r_{\rm mit} \approx 70$ indicates the ``true'' metal-insulator transition due to interstitial self-doping proposed in this paper, and is distinct from $r_{\rm melt}$ below which the crystalline order vanishes.
    $r_{\rm melt}$ is expected to be smaller than the value for a direct FL--WC transition from quantum Monte Carlo calculations, $r_{\rm melt}^* \approx 31$ \cite{drummond2009QMC}, due to the existence of the intermediate MeC phase.  (Additional microemulsion phases may be expected~\cite{spivak2006transport} as well, especially for $r_s \sim r_{\rm melt}^*$.)
    See Sec.~\ref{sec:an intermediate Phase of 2DEG} for a detailed discussion of the conjectured phase diagram.
   }
    \label{fig:phase diagram}
\end{figure}

Apart from the charge ordering, there is another subtle issue regarding the magnetism.
In the FL regime, the paramagnetic state seems to be most favored \cite{drummond2009QMC}.
Deep within the WC phase, the magnetism is determined by various ring-exchange processes.
The exchange coefficients can be calculated using the semi-classical instanton approximation \cite{Roger1984WKB,roger1983RMP, chakravarty1999WC, Katano2000WKB, voelker2001disorder}, the validity of which is well-tested by a numerically exact path integral Monte Carlo calculation \cite{Ceperley2001exchange}.
These calculations show that the WC is a ferromagnet for large enough $r_s > r_{\rm F}^{\rm wc} \approx 175$ \cite{Ceperley2001exchange} and a (highly frustrated) antiferromagnet \cite{chakravarty1999WC} below $r_{\rm F}^{\rm wc}$ (Fig.~\ref{fig:phase diagram}).
However, the predicted energy scale for the ring-exchange processes within the WC phase is too small to account for the typical magnetic energy scale of the insulating  phase observed in the large $r_s$ regime of various 2DEG systems \cite{hossain2020ferromagnetism, kim2021discovery, falson2022competing}.
This prompted some of the present authors to propose a kinetic mechanism that accounts for higher-temperature magnetism in such a phase mediated by interstitial hopping processes \cite{kim2022interstitial} 
\footnote{There is a sign error in the correlated hopping terms $t_2$ and $t_2'$ in Eq.~(4) of Ref.~\cite{kim2022interstitial}, which led the authors to erroneously claim that the interstitial dynamics induces a fully-polarized ferromagnet. This is corrected as in Eq.~\eqref{eq:interstitial Hamiltonian} of the current paper. 
The resulting magnetism due to the interstitial dynamics is more complicated as discussed in Sec.~\ref{sec:Kinetic magnetism}.
}.

\begin{figure}[t]
    \centering
    \includegraphics[width = 0.48\textwidth]{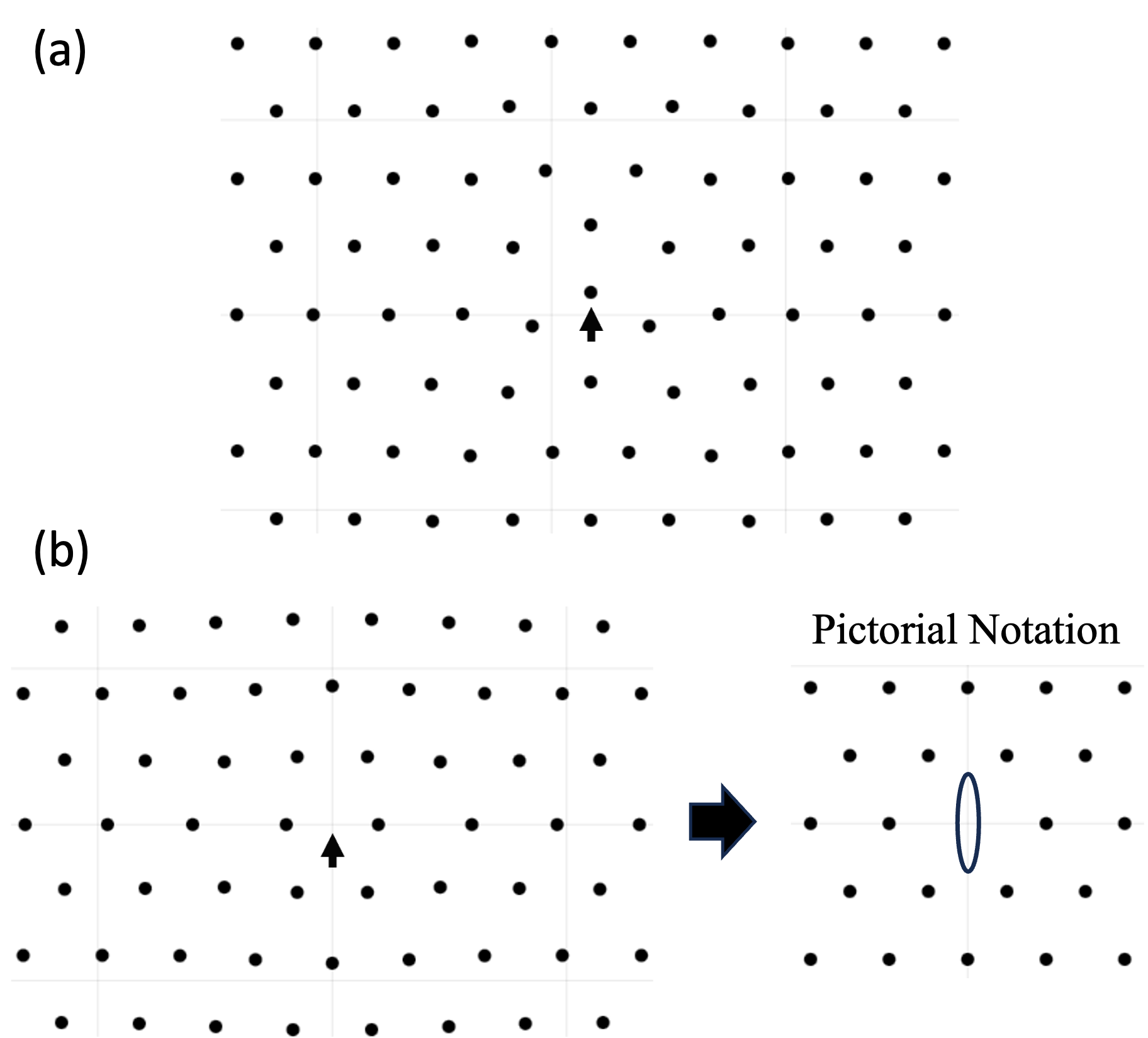}
    \caption{(a) A classical centered interstitial and (b) a vacancy configuration.
    Small black arrows are drawn to indicate the positions of the interstitial (panel a) and the vacancy (left panel of b).
    The vacancy configuration has $D_2$ symmetry, and not the full $D_6$ symmetry of the underlying WC; therefore, {a vacancy has three possible} orientations $\alpha$. 
    We introduce a pictorial notation for the vacancy for later convenience.}
    \label{fig:defect configurations}
\end{figure}

In this paper, using the semi-classical  instanton approximation, we carry out a comprehensive study  of the quantum dynamics of an interstitial and a vacancy defect (Fig.~\ref{fig:defect configurations}), two point defects of a WC with the smallest classical creation energies \cite{Fisher1979defect,Cockayne1991defect}
\footnote{Note that the energetics of these point defects are also studied using the path integral Monte Carlo method in Ref.~\cite{Ceperley200defect}, but the contributions from defect hopping and exchange processes are ignored.}.
We first review the formulation of the standard instanton technique and apply it to derive effective Hamiltonians describing various exchange and defect hopping processes illustrated in Fig.~\ref{fig:Tunneling} (Sec.~\ref{sec:The Semi-classical approximation}).
In Sec.~\ref{sec:Energy of WC defects and the self-doping instability of a WC}, we calculate the energy of an interstitial and a vacancy via finite-size exact diagonalization of the derived effective Hamiltonians.
Interestingly, the resulting semi-classical expression for the interstitial energy, when extrapolated to a large but finite $r_s$, vanishes around $r_s = r_{\rm mit} \approx 70$, signaling a possible self-doping instability to a partially melted WC below $r_{\rm mit}$.
From this, we propose the existence of a metallic electron crystal (MeC) phase as an intermediate phase of the 2DEG (Sec.~\ref{sec:an intermediate Phase of 2DEG}).
In Sec.~\ref{sec:Kinetic magnetism}, we discuss the magnetic correlations induced by interstitial and vacancy hopping processes. 
Such kinetic processes induce magnetism with much higher energy scales than the ring-exchange processes of the pure WC;
this could be experimentally probed by controlled doping of a WC that is commensurately locked to a weak periodic substrate potential.
Our principal results are summarized in Figure~\ref{fig:phase diagram}.
We conclude with a remark on the fate of the phase diagram in the presence of weak quenched disorder in Sec.~\ref{sec:Effects of weak disorder}.

\section{The Semi-classical approximation}
\label{sec:The Semi-classical approximation}
We first review the standard semi-classical instanton method as applied to the ideal 2DEG (\ref{eq:2DEG}) in the large $r_s$ limit. 
The exact partition function of the (fermionic) 2DEG  is 
\begin{align}
    &Z = \int d^{2N} \v r_0 \sum_{P \in S_N} \frac{(-1)^P}{N!} \sum_{\spin} 
    \left <P \v r_0, P\spin  \right| e^{-\beta H} \left | \v r_0, \spin  \right>, \label{eq:partition function}
    \\
    & \left <P \v r_0, P\spin  \right| e^{-\beta H} \left | \v r_0, \spin  \right>  = \delta_{\spin, P\spin} \left <P \v r_0  \right| e^{-\beta H} \left | \v r_0 \right>,
    \\
    &
    \left < 
    \v r_0'
      \right| e^{-\beta H} \left | \v r_0  \right> 
    =
    \int_{\tilde{\v r}(0) = \tilde{\v r}_0}^{\tilde{\v r}(\tilde \beta) = 
    \tilde {\v r}_0'} D\tilde {\v r}(\tau) e^{- \sqrt{r_s}S},
    \\
    &
    \label{eq:action}
    S = \int_0^{\tilde \beta} d\tau \left [ \frac 1 2   \left (\frac{d{\tilde{\v r}}}{d\tau} \right)^2 + V (\tilde{\v r} ) - V_0 \right ], 
    \\
    &
    V(\v r) \equiv \sum_{i<j} \frac{1}{|\v r_i -\v r_j|},
\end{align}
where $\v r(\tau) \equiv \{\v r_i (\tau) \}$ are the positions of $N$ electrons in imaginary time, $\v r_0 \equiv \{\v r_i(\tau =0)\}$ are their initial positions,  $\spin \equiv \{ \sigma_i = \, \uparrow,\downarrow\}$ are their respective spin indices,  and $\beta = 1/k_B T$ is the inverse temperature.
The sum over $N!$ permutations, $P,$ of the coordinates and the sign factor $(-1)^P$ encode the  fermionic exchange statistics. 
For bosonic particles, one should merely substitute $(-1)^P \rightarrow +1$.
The 2DEG Hamiltonian~(\ref{eq:2DEG}) does not act on the electron spins,  hence the $\delta_{\spin, P\spin}$ factor in the second line above.
The third and fourth lines are the path integral representation of the $N$-electron propagator.
The action is rescaled to make the $r_s$ dependence manifest by introducing 
dimensionless coordinates, $\tilde {\v r} \equiv {\v r}/a_0$, and dimensionless imaginary time, $\tau$.
Correspondingly, $\tilde \beta \equiv \beta E^*$ is a dimensionless inverse temperature, where 
$E^* \equiv e^2/(4 \pi \epsilon a_{\rm B} r_s^{3/2})$.
The path integral measure is also defined as an integration over the dimensionless coordinate $\tilde{\v r}(\tau)$.
The minimum potential energy $V_0 = {\rm min}_{\tilde{\v r}}V(\tilde{\v r})$ is subtracted for later convenience~%
\footnote{When considering exchange processes in a pure WC, $V_0$ corresponds to the classical WC energy. For tunneling processes involving a defect, $V_0$ corresponds to the classical energy of the defect.}.
The Coulomb interaction (last line) is computed numerically using the standard Ewald method.
As usual, the presence of a uniform neutralizing positively-charged background is assumed. 
Henceforth, we will drop tildes from the rescaled coordinates to simplify notation: $\tilde {\v r } \to \v r.$ 
We focus on the zero temperature phase of the problem, and hence will always take $\beta \to \infty$ in the end.

We approach this problem using a semi-classical instanton approximation, 
which is asymptotically exact in the $r_s \to \infty$ (strong coupling) limit.
In Sec.~\ref{sec:WC ring-exchange processes}, we briefly review the semi-classical derivation of ring-exchange processes in the WC.
In Sec.~\ref{sec:Processes involving a single interstitial} and \ref{sec:Processes involving a single vacancy}, we consider tunneling processes involving a single interstitial and vacancy, respectively, and derive the corresponding effective Hamiltonians describing their dynamics.
The application of the semi-classics to a bosonic system is addressed in Sec.~\ref{sec:Two-Dimensional Bose Gas}.

\subsection{Wigner crystal ring-exchange processes}
\label{sec:WC ring-exchange processes}

In the $r_s\to \infty$ limit, the classical ground state manifold consists of a triangular lattice WC with $2^N$-fold degeneracy in spin states.
The lifting of this degeneracy and the nature of the resulting magnetic order is determined for $1 \ll r_s < \infty$ by WC ring-exchange processes.  
Various ring-exchange processes correspond to distinct instanton solutions of the action and 
can be calculated via the dilute instanton approximation \cite{thouless1965, roger1983RMP, Roger1984WKB, chakravarty1999WC, Katano2000WKB, voelker2001disorder, kim2022interstitial}, which we briefly review below. (See Refs.~\cite{Katano2000WKB, voelker2001disorder} for more details.)  
The result is an effective spin Hamiltonian expressed as a sum over all ring-exchange processes:
\begin{align}
\label{eq:WC exchanges}
    H_{\mathrm{eff}}^{\mathrm{wc}} = -\sum_{a} (-1)^{P_a}\, {J_a} \, \big( \mathcal{\hat P}_a +\mathcal{\hat P}^{-1}_a \big) ,
\end{align}
where the semi-classical calculation gives a leading-order large $r_s$ asymptotic expression for $J_a$.
Here, $\mathcal{\hat P}_a$ is the permutation operator corresponding to the permutation $P_a$, and can be decomposed as a product of two-particle exchange operators.
The two-particle exchange operators, in turn, can be written in terms of spin operators as $\mathcal {\hat P}_{(i,j)} = 2(\vec{S}_i \cdot \vec{S}_j + \frac{1}{4})$.

To illustrate how this works, recall the familiar problem of the semi-classical calculation of the tunnel splitting in a symmetric  double-well potential \cite{coleman1988aspects, altlandSimons2010,zinnJustinQFT}.  For large enough $\beta$ such that $\beta \hbar \omega_0 \gg 1$, the excited states in each well can be neglected.  (Here, $\omega_0$ is the  oscillation frequency in either well.)  
In this limit, we obtain asymptotic relations
\begin{align}
\label{diag}
&\left < \v r_0  \right| e^{-\beta H} \left | \v r_0 \right> 
\sim |\psi(\v r_0)|^2  e^{-\beta E_0} \cosh(\beta \Delta) , \\*
&\left < -\v r_0  \right| e^{-\beta H} \left | \v r_0 \right> 
\sim |\psi(\v r_0)|^2  e^{-\beta E_0} \sinh(\beta \Delta) ,
\label{offdiag}
\end{align}
where the minima of the two wells are at $\pm \v r_0$, $|\psi(\v r_0)|^2=|\psi(-\v r_0)|^2$ is the probability density of the wave function at these positions, and $E_0$ and $2\Delta$ are, respectively, the mean energy and the splitting between the even and odd parity ground states.  
The right-hand side of each expression is obtained by inserting the resolution of the identity on the left-hand side.

\begin{figure}[t]
    \centering
\includegraphics[width = 0.48\textwidth]{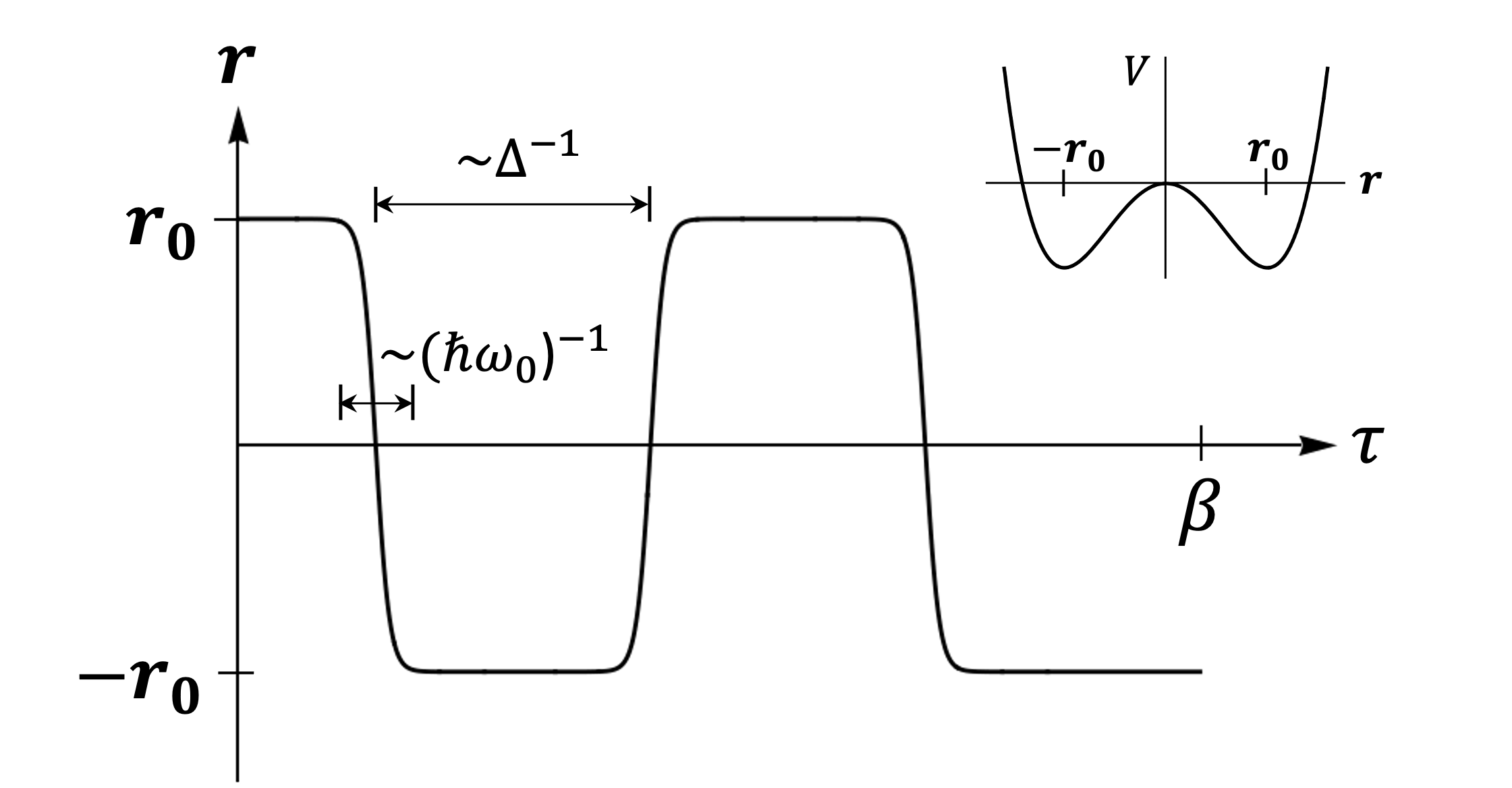}
    \caption{An example of a multi-instanton configuration for the double well potential shown in the inset. The ``size" of each instanton in imaginary time is $\sim 1/ \hbar \omega_0$ and the ``distance"  between them is $\sim 1/ \Delta$.}
    \label{fig:instanton}
\end{figure}

In viewing this same problem from the path integral perspective in the semi-classical limit, one first solves for the instanton path---the smallest action path that begins at the bottom of one well and ends at the bottom of the other. 
The net duration (in imaginary time) of this tunneling event is of order $\omega_0^{-1}$.  
We then sum over multiple such instanton events to obtain an expression of the same form as above, where the diagonal (off-diagonal) propagator in Eq.~\ref{diag} (Eq.~\ref{offdiag}) contains all the terms with an even (odd) number of events.  
Expanding these expressions in power series, one sees that the typical number of tunneling events is $\sim \beta \Delta$ and the mean imaginary time interval between them is of order $\hbar/\Delta$.
Note that in the semi-classical limit $\hbar/\Delta \gg \omega_0^{-1}$, the instantons are dilute and hence effectively non-interacting (see Fig.~\ref{fig:instanton}).  
Looked at another way, for a range of temperature such that $\hbar \omega_0 \gg T \gg \Delta$, where multiple instanton events can be neglected, we can compute $\Delta$ as
\begin{align}
\label{eq:Delta}
    \Delta = \beta^{-1} \, \frac {\left < -\v r_0  \right| e^{-\beta H} \left | \v r_0 \right>\vert_{\text{1-inst}}} {\left < \v r_0  \right| e^{-\beta H} \left | \v r_0 \right>\vert_{\text{0-inst}}} ,
\end{align}
where the subscripts designate the number of instanton events.

\begin{figure*}[t]
    \centering
\includegraphics[width = \textwidth]{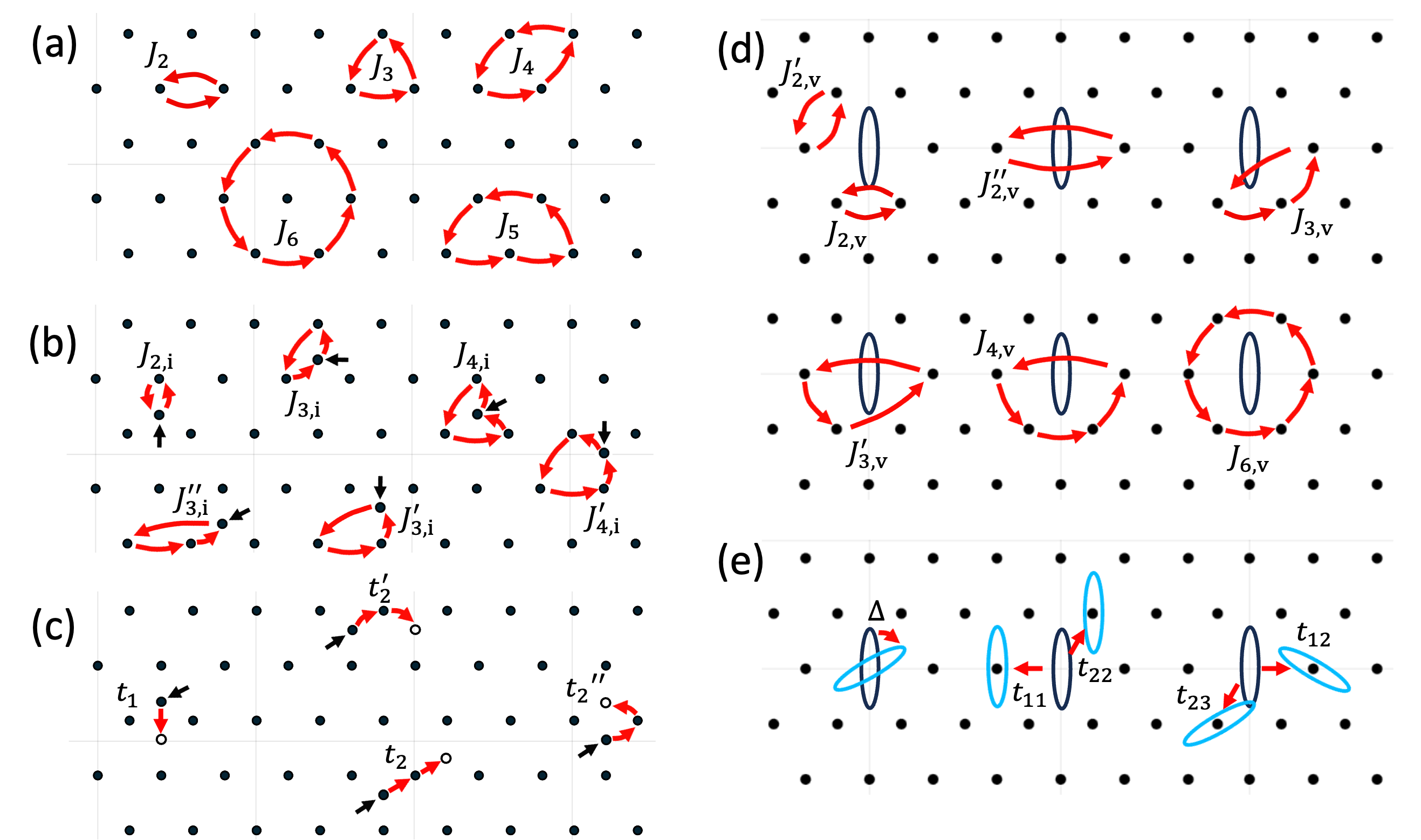}
    \caption{Tunneling processes considered in this paper. (a) WC exchange processes. (b) Exchange processes involving an interstitial. (c) Interstitial hopping processes. (d) Exchange processes involving a vacancy. (e) Vacancy hopping processes.
    In  (b,c), black arrows indicate the positions of interstitials.
    In (e), a black (cyan) oval denotes an initial (final) vacancy configuration corresponding to each vacancy hopping process. 
    $t_{11},t_{12},t_{22},t_{23}$ exhaust all the nearest-neighbor vacancy hopping processes; others are related to one of these by symmetry.
    Panels (a-c) are adapted from Ref.~\cite{kim2022interstitial}.}
    \label{fig:Tunneling}
\end{figure*}

The analysis is somewhat more complicated but structurally similar for the present problem.  Consider the propagator $\left <P_a \v r_0  \right| e^{-\beta H} \left | \v r_0 \right>$  where $\v r_0$ is  an initial WC configuration and $P_a$  is the permutation corresponding to a particular ring-exchange process [see Fig.~\ref{fig:Tunneling}(a)].
In the semi-classical (large $r_s$) limit, this propagator is again expressible as a weighted sum over multi-instanton {contributions.}
For temperatures such that $\hbar\Omega \gg T \gg J_a$, where $J_a$ is the tunnel splitting corresponding to the process $P_a$, the propagator is dominated (up to symmetry) by a single ``$a$''  instanton contribution associated with the path
$\v r^{(a)}(\tau)$ with the smallest action subject to the boundary conditions $\v r^{(a)}(0) = \v r_0$ and $\v r^{(a)}(\tilde \beta) = P_a \v r_0$.  
Here, $\hbar\Omega/2 \sim r_s^{-3/2}$ is the zero-point energy of the WC, while $J_a$ is exponentially small in $\sqrt{r_s}$ at large $r_s$. 
The single-$a$-instanton contribution to the propagator can be expressed as
\begin{align}
    &\left <P_a \v r_0 \right| e^{-\beta H} \left | \v r_0 \right> \vert_{a,\text{1-inst}}
    \nonumber \\*
    &\ \approx e^{-\sqrt{r_s} S_a} \int_{\delta \v r(0)= \v 0}^{\delta \v r(\tilde \beta )= \v 0} D \delta \v r(\tau) \, e^{-\frac{1}{2} \sqrt{r_s} \int_0^{\tilde \beta } \delta \v r(\tau)^T {\bf {\hat { M}}}^{(a)}(\tau) \delta \v r(\tau)}
    \nonumber \\*
    &\ = e^{-\sqrt{r_s} S_a} \left ( \det\big [\sqrt{r_s} \ {\bf {\hat { M}}}^{(a)}(\tau) \big ] \right )^{-1/2},
    \\
    &\hat M^{(a)}_{ij}(\tau) 
    \equiv \frac{\delta^2 S}{\delta r^{(a)}_i(\tau) \, \delta r^{(a)}_j(\tau)} 
    = -\delta_{ij}\frac{\partial^2}{\partial \tau^2} + \partial_i \partial_j V\big [\v r^{(a)}(\tau) \big ], 
    \label{eq:M elements}
\end{align}
%
%
%
where $S_a \equiv S\big [\v r^{(a)}(\tau) \big ]$ with the trajectory $\v r^{(a)}(\tau)$ satisfying $\delta S\big [\v r^{(a)}(\tau)\big ] = 0$, and
$\delta \v r(\tau) \equiv  \v r(\tau) -  \v r^{(a)}(\tau)$ is the fluctuation coordinate.
Fluctuations are treated within a harmonic approximation around the semi-classical path.
In Eq.~\eqref{eq:M elements}, the derivative $\partial_i$ is with respect to the normalized coordinates.
Note that ${\bf \hat M}^{(a)} $ has a zero eigenvalue solution $\dot{\v r}^{(a)}(\tau)$ corresponding to the translation in imaginary time, which has to be treated with care \cite{altlandSimons2010, coleman1988aspects, zinnJustinQFT,voelker2001disorder,Katano2000WKB}.
Separating the zero mode contribution from the full determinant, one obtains 
\begin{align}
\label{eq:1 instanton}
    &\left <P_a \v r_0  \right| e^{-\beta H} \left | \v r_0 \right> \vert_{a,\text{1-inst}
    } 
    \\*
    &=
    \beta \, \frac{e^2}{4\pi \epsilon a_{\rm B} r_s^{3/2}}
    \sqrt{\frac{S_a}{2\pi}}\cdot   e^{-\sqrt{r_s} S_a} \left ( {\rm det}' \big [ \sqrt{r_s} \, {\bf \hat M}^{(a)}(\tau) \big ] \right )^{-\frac 1 2}, \nonumber
\end{align}
where the prime denotes that the zero eigenvalue must be omitted in the calculation of the determinant.
Note that since an instanton is a localized object with a characteristic size $\Delta \tau = 
\Omega^{-1}$, one can neglect the exponentially small correction from its tail  
provided $\beta \hbar\Omega \gg 1$.

\begin{figure}[t]
\includegraphics[width = 0.48\textwidth]{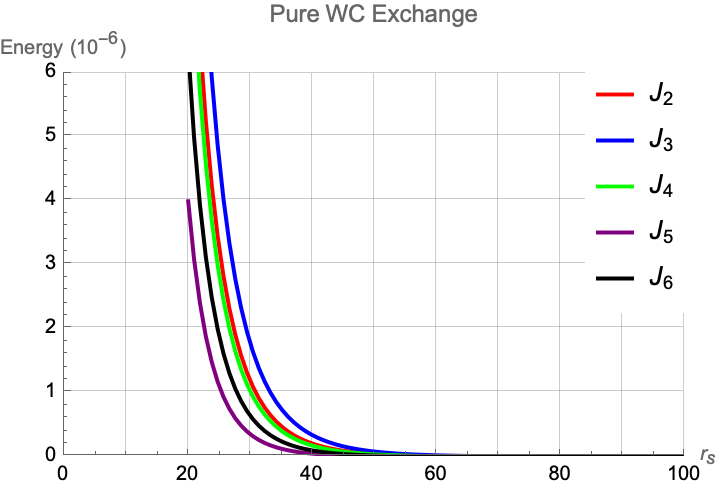}
    \caption{Exchange coefficients of the pure WC (in units of the Hartree energy $e^2/4\pi \epsilon a_B$) as a function of $r_s$, calculated from the semi-classical expression~(\ref{eq:tunnel splitting}).
    The processes corresponding to $J_2$,\dots, $J_6$ are schematically illustrated in Fig.~\ref{fig:Tunneling}(a).
    Within the WC phase ($r_s \gtrsim 30$),  the instanton approximation is well-justified for the calculation of these ring-exchange processes since $\sqrt{r_s} S_a \gg 1$. 
    }
    \label{fig:pureWCexchange}
\end{figure}

On the other hand, the diagonal propagator in the zero instanton sector $\left <\v r_0  \right| e^{-\beta H} \left | \v r_0 \right> \vert_{\text{0-inst}}$ can be obtained by making a harmonic approximation of $V$ around $\v r_0$:
\begin{align}
\label{eq:0 instanton}
    &\left <\v r_0  \right| e^{-\beta H} \left | \v r_0 \right> \vert_{\text{0-inst}
    } \approx  \left ( \det\big [ \sqrt{r_s} \, {\bf \hat M}^{(0)}(\tau) \big ] \right )^{-\frac 1 2} ,
    \\
& 
\  
{\hat M}^{(0)}(\tau) \equiv  -\delta_{ij}\frac{\partial^2}{\partial \tau^2} + \partial_i \partial_j V(\v r_0 ).
\end{align}
Normalizing the propagator in the one instanton sector by that in the zero instanton sector, as in Eq.~\eqref{eq:Delta}, one obtains
\begin{align}
    J_a &= \beta^{-1} \, \frac {\left <P_a \v r_0  \right| e^{-\beta H} \left | \v r_0 \right>\vert_{a,\text{1-inst}}} {\left < \v r_0  \right| e^{-\beta H} \left | \v r_0 \right>\vert_{\text{0-inst}}} 
    \nonumber \\*
    \label{eq:tunnel splitting}
    &= 
    \frac{e^2}{4\pi\epsilon a_{\rm B}}
    \cdot \frac {{A_a}} {{ r_s^{5/4} }}\sqrt{\frac{S_a}{2\pi}}\ e^{-\sqrt{r_s}S_a} >0,
     \\*
    A_a &= \left [ \frac{ {\rm det}'\big (-\partial_\tau^2 + V''\big [\v r^{(a)}(\tau)\big ]\big )}{\det\big (-\partial_\tau^2 + V''(\v r_0) \big )} \right ]^{-\frac 1 2}, \label{eq:fluctuation determinant}
\end{align}
where $A_a$ is called a ``fluctuation determinant,'' 
calculated in the normalized coordinates with $r_s = 1$, and the $ \beta \to \infty$ limit is implicitly taken in the end. 
In the second line, the extra factor of $r_s^{1/4}$ comes from the normalization of the determinant
\begin{align}
    \left (\frac{  {\rm det}'\big [ \sqrt{r_s} \, {\bf \hat M}^{(a)}(\tau) \big ]}{\det\big [ \sqrt{r_s} \, {\bf \hat M}^{(0)}(\tau) \big ]} \right )^{-\frac 1 2} = r_s^{1/4}   \left (\frac{  {\rm det}'\big [ {\bf \hat M}^{(a)}(\tau) \big ]}{\det\big [ {\bf \hat M}^{(0)}(\tau) \big ]} \right )^{-\frac 1 2}. \nonumber 
\end{align}
Hence, $A_a$ (\ref{eq:fluctuation determinant})  (and also $S_a$) are dimensionless numbers with no $r_s$ dependence.
In Eq.~\eqref{eq:fluctuation determinant},  $V''$ denotes the hessian matrix of $V$.
We refer readers to Appendix \ref{app:Details of the numerical  calculation of $S_a$ and $A_a$} for the details of the numerical calculation of $S_a$ and $A_a$.
For the ring-exchange processes illustrated in Fig.~\ref{fig:Tunneling}(a), we quote the results for $S_a$ and $A_a$ from Ref.~\cite{voelker2001disorder}: 
$S_2=1.64$, $A_2 = 1.30$; $S_3=1.53$, $A_3=1.10$; $S_4=1.66$, $A_4=1.24$; $S_5=1.91$, $A_5=1.57$; $S_6=1.78$, $A_6= 1.45$.
Our calculations, and those of Ref.~\cite{Katano2000WKB}, agree with these values.
The resulting exchange coefficients calculated from the semi-classical expression \eqref{eq:tunnel splitting} are shown in Fig.~\ref{fig:pureWCexchange}.

The remaining issue concerns the sign factor $(-1)^{P_a}$ that enters $H_{\rm eff}^{\rm wc}$ in Eq.~\eqref{eq:WC exchanges}, which is due to the anti-symmetry of the many-body electronic wave function (see chapter V of Ref.~\cite{roger1983RMP} for an explanation).
As recognized by Thouless~\cite{thouless1965}, this implies that a ring-exchange process involving an even (odd) number of electrons mediates an antiferromagnetic (ferromagnetic) interaction.

\subsection{Processes involving a single interstitial}
\label{sec:Processes involving a single interstitial}

\begin{table}[t]
\includegraphics[width = 0.48\textwidth]{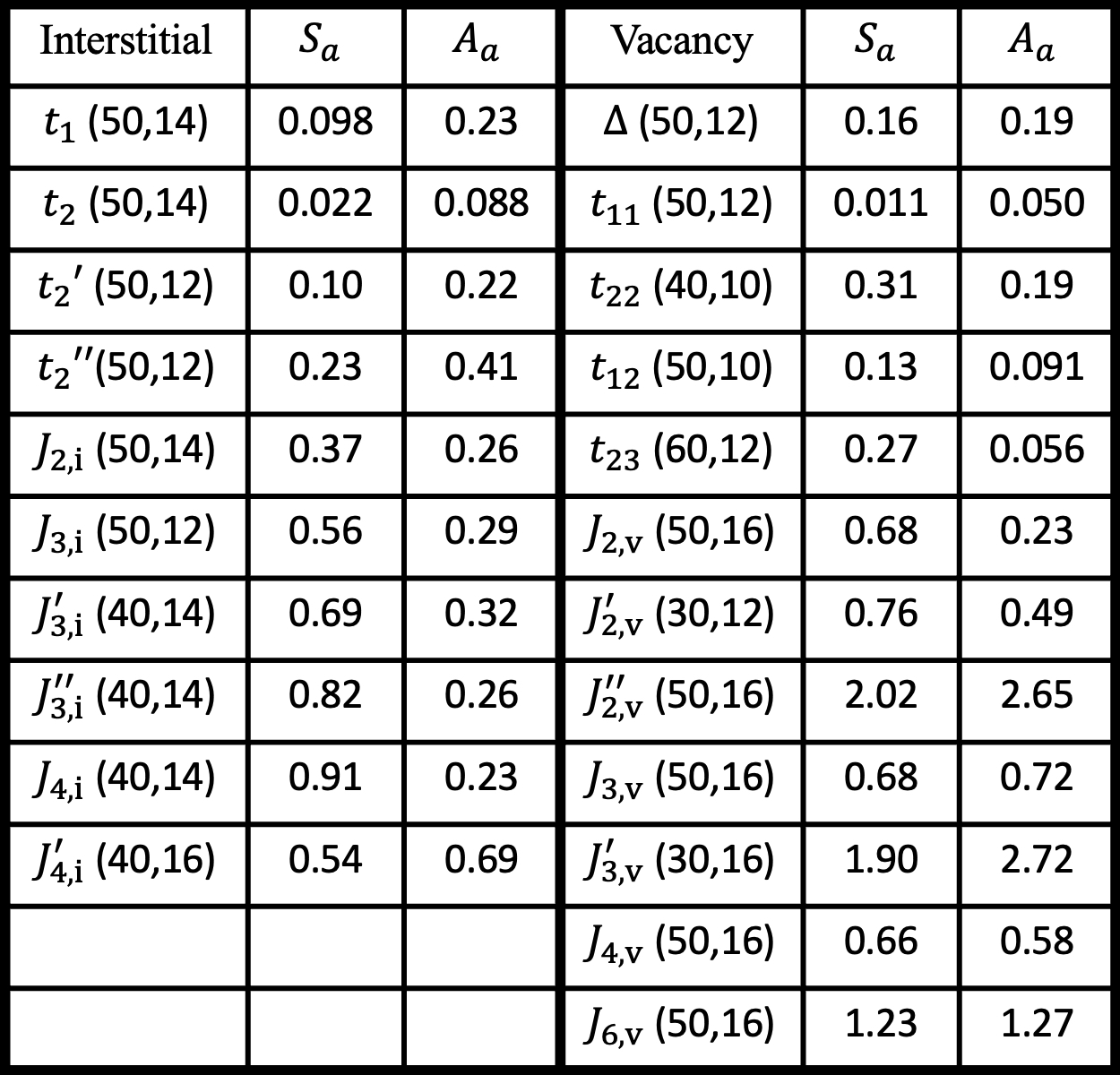}
    \caption{Dimensionless actions $S_a$ and fluctuation determinants $A_a$ for tunneling processes illustrated in Fig.~\ref{fig:Tunneling}(b-e) calculated in this paper. 
    The parentheses in the first and fourth columns denote $(N_{\rm move}, M)$, where $N_{\rm move}$ is the number of electrons that are allowed to adjust their positions during minimization and $M$ is the number of time slices for the discretized tunneling paths (i.e., there are $M-1$ intermediate configurations).
    Processes for a centered interstitial (vacancy) are calculated in a hexagonal supercell with $12 \times 12+ 1$ ($10\times 10 -1$) electrons starting and ending at fully relaxed defect configurations. }
    \label{table:results}
\end{table}

\begin{figure*}[t]
    \centering
    \hspace{-6.5mm}
\includegraphics[width = 0.9 \textwidth]{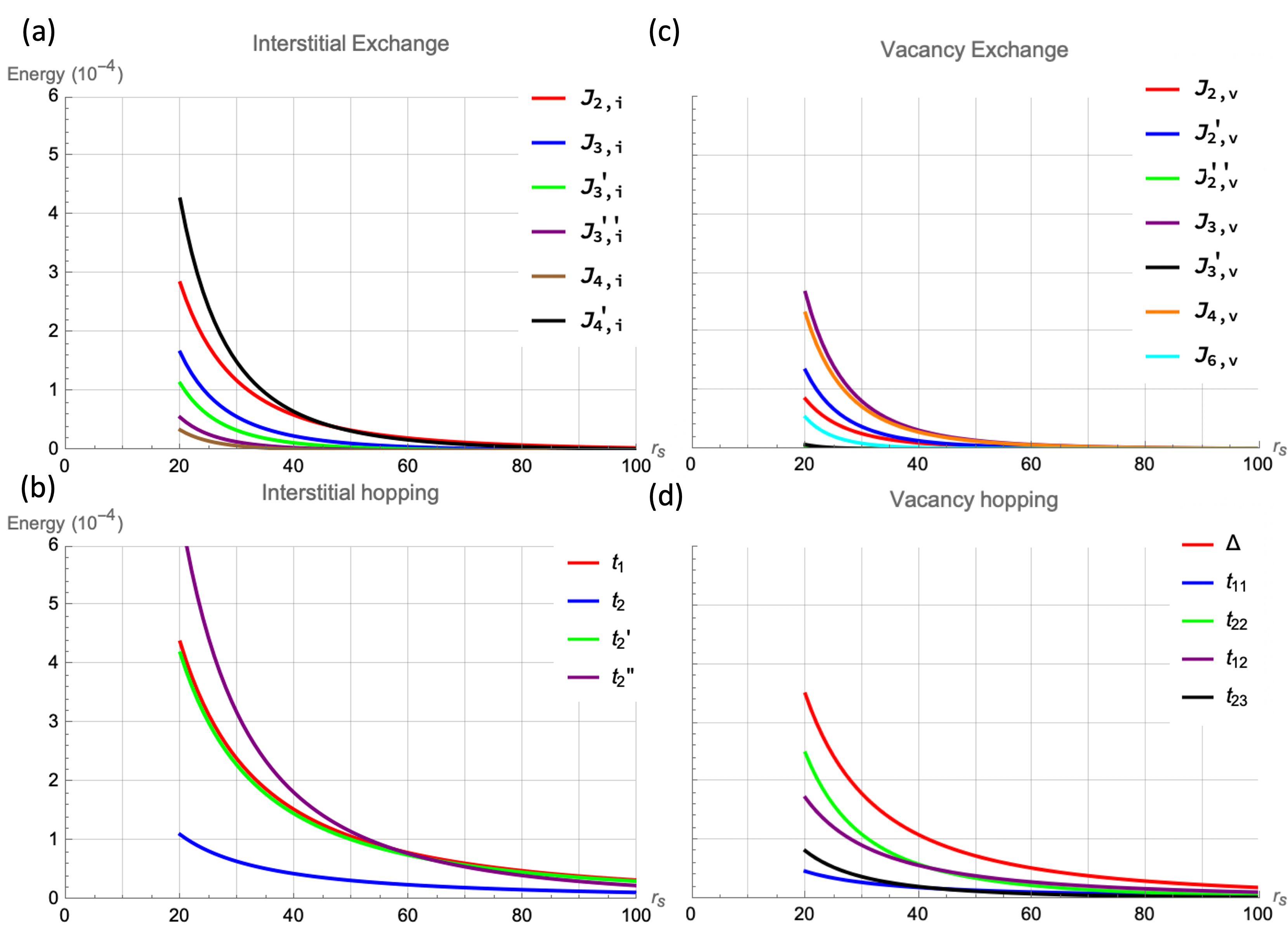}
    \caption{Hopping matrix elements and exchange coefficients involving a defect (in units of ${e^2}/{4 \pi \epsilon a_B}$) within the semi-classical approximation. 
    Note that the 
    y-axis scale here is a factor of 100 larger than in Fig.~\ref{fig:pureWCexchange}. 
    Hence, the dynamical processes involving an interstitial or a vacancy have much larger energy scales than the 
    exchange processes in the pure WC. 
    }
    \label{fig:matrix_elements}
\end{figure*}

Tunneling processes involving a single centered interstitial (CI) defect [Fig.~\ref{fig:defect configurations}(a)] were first considered in Ref.~\cite{kim2022interstitial}.
We correct and refine the results obtained there:
(1) The sign error in the correlated hopping terms $t_2$ and $t_2'$ in Eq.~(4) of Ref.~\cite{kim2022interstitial} is corrected in Eq.~\eqref{eq:interstitial Hamiltonian};
(2) We improve the estimate of the classical action (which is done by solving the classical equations of motion for a finite sized system with periodic boundary conditions) using a hexagonal, instead of a rectangular, supercell with $12\times 12 +1$ electrons;  
and (3) We explicitly calculate the fluctuation determinants $A_a$ rather than simply making dimensional estimates. 
Fig.~\ref{fig:Tunneling}(b-c) show the tunneling processes considered in this paper with the corresponding $S_a$ and $A_a$ listed in Table~\ref{table:results}.
The hopping matrix elements $t_a>0$ are again expressed in terms of $S_a$ and $A_a$ as in Eq.~\ref{eq:tunnel splitting}.
Note that four hopping processes have smaller actions than those of exchange processes and hence are more important when $r_s \gg 1$.
The effective Hamiltonian in the presence of a dilute concentration of interstitials is (corrected from Ref.~\cite{kim2022interstitial})
\begin{align}\label{eq:interstitial Hamiltonian}
    H_{\mathrm{eff}}^{\rm i} =
    &-t_1 \sum_{\left<n,n^\prime\right>}
    \sum_{\sigma} \,
    c^{\dagger}_{n,\sigma}c_{n^\prime,\sigma} 
    \nonumber \\*
    &-t_2 \sum_{\substack{(n,j,n^\prime) \\ \in (t_2 \: \mathrm{path})}} 
    \sum_{\sigma,\sigma'} \,
     f^{\dagger}_{j,\sigma} c^{\dagger}_{n,\sigma'} f_{j,\sigma'} c_{n^\prime,\sigma} 
    \nonumber \\*
    &- t_2' \sum_{\substack{(n,j,n^\prime) \\ \in (t_2' \: \mathrm{path})}}
    \sum_{\sigma,\sigma'} \,
    f^{\dagger}_{j,\sigma} c^{\dagger}_{n,\sigma'}  f_{j,\sigma'} c_{n^\prime,\sigma}
    \nonumber \\*
    &- t_2'' \sum_{\substack{(n,j,n^\prime) \\ \in (t_2'' \: \mathrm{path})}}
    \sum_{\sigma,\sigma'} \,
     f^{\dagger}_{j,\sigma} c^{\dagger}_{n,\sigma'} f_{j,\sigma'} c_{n^\prime,\sigma}
    \nonumber \\*
    &- \sum_{a \in (\text{CI ex.})} (-1)^{P_{a,{\rm i}}} J_{a,{\rm i}} \, 
    \big( \mathcal{\hat P}_{a,{\rm i}} + \mathcal{\hat P}_{a,{\rm i}}^{-1} \big)
    \nonumber \\*
    &+ \ \cdots \ + \ \left[ U=\infty \right] ,
\end{align}
%
%
%
%
%
where $f_{j\sigma}^\dagger$ ($c_{n,\sigma}^\dagger$) is the creation operator of electrons that live on the  WC sites $j$ (triangular plaquette centers $n$) and
the $U=\infty$ condition precludes any double occupancy. 
$\sigma,\sigma' = \ \uparrow, \downarrow$ are the spin indices that are summed over.
$a \in (\rm{CI\ ex.})$ denotes one of the exchange processes involving an interstitial shown in Fig.~\ref{fig:Tunneling}(b).
The omitted terms correspond to hopping and exchange processes other than those shown in Fig.~\ref{fig:Tunneling}(b-c) and direct (elastic) interactions between interstitials \cite{footnote:interaction}.
Figure~\ref{fig:matrix_elements}(a-b) shows the hopping matrix elements ($t$) and exchange coefficients ($J$) for processes involving an interstitial calculated from the semi-classical expression~\eqref{eq:tunnel splitting}.

\subsection{Processes involving a single vacancy}
\label{sec:Processes involving a single vacancy}

The classical vacancy defect has $D_2$ symmetry instead of the full $D_6$ symmetry of the underlying triangular lattice \cite{Cockayne1991defect} [see Fig.~\ref{fig:defect configurations}(b)].
Therefore, associated with each location of a vacancy, there are 3 inequivalent orientations related by $C_6$ rotations.
We will denote 
{these by an index} $\alpha = 1, 2, 3$; $ \alpha = 2$ and $3$ are related to $\alpha=1$ by $\mathcal C_6$ and $\mathcal C_6^2$ respectively. 

We considered tunneling processes involving a single vacancy defect as illustrated in Fig.~\ref{fig:Tunneling}(d-e), with their corresponding values of $S_a$ and $A_a$  listed in Table~\ref{table:results}.
The calculation is done in a hexagonal supercell containing $10 \times 10 -1$ electrons.
Again, matrix elements $\Delta, t_a, J_{a,{\rm v}}>0$ are given by Eq.~\eqref{eq:tunnel splitting} in terms of $S_a$ and $A_a$.
Note that, as in the interstitial case, the tunnel barriers ({determined by} $S_a$) for hopping processes are smaller than those for exchange processes
\footnote{Note that the tunnel barrier for $t_{11}$ is anomalously smaller than that for any other processes, implying that the dynamics of the vacancy is predominantly uni-directional in the $r_s \to \infty$ limit.
Such a peculiar defect dynamics has also been predicted to occur in solid helium \cite{andreev1976diffusion}.
Such a ``restricted mobility'' would in turn lead to an exponentially slow (in $\sqrt{r_s}$) thermalization in the presence of a dilute vacancy concentration.
However, within the semi-classical approximation, $t_{11}$ is the largest energy scale only when $r_s \gtrsim 320$, and this peculiar feature may be difficult to observe in practice.
}.

The resulting effective Hamiltonian describing the dynamics of vacancies can be written straightforwardly as follows. 
First, corresponding to each orientation $\alpha$ of a vacancy, we introduce a (hard-core) bosonic operator $b_{i,\alpha}^{\dagger}$ that suitably relaxes the positions of the WC electrons near the vacancy site $i$ to the associated configuration that minimizes the (classical) Coulomb energy.
The operator that creates a vacancy in the WC at site $i$ with orientation $\alpha$ is thus $f_{i,\sigma}b^{\dagger}_{i,\alpha}$.
Then, with the definitions $\v b_i^{\dagger} \equiv [b_{i,1}^{\dagger}, b_{i,2}^{\dagger}, b_{i,3}^{\dagger}]$ and 
\begin{align}
    \mathfrak{D} \equiv 
    \begin{bmatrix}
   0 & \Delta & \Delta \\  \Delta & 0 & \Delta \\ \Delta & \Delta & 0
    \end{bmatrix},\ 
    \Upsilon \equiv \begin{bmatrix}
   t_{11} & t_{12} & t_{12} \\  t_{12} & t_{22} & t_{23} \\ t_{12} & t_{23} & t_{22}
    \end{bmatrix},\ 
    \mathcal C_6  = \begin{bmatrix}
        0 & 1 & 0 \\ 0 & 0 & 1\\ 1& 0 &0
    \end{bmatrix},
\end{align}
the effective Hamiltonian in the presence of a dilute concentration of vacancies is
\begin{align}\label{eq:vacancy Hamiltonian}
    H_{\rm eff}^{\rm v} = 
    &-\sum_{i,\sigma} \Bigg[ \, \v b^{\dagger}_i \mathfrak{D} \v b_i
    + \sum_{\delta = \pm \v e_1} f^{\dagger}_{i,\sigma} f_{i+\delta, \sigma} \v b^{\dagger}_{i+\delta} \Upsilon\v \, b_{i} 
    \nonumber \\*
    &\qquad\quad +\sum_{\delta = \pm \v e_2} f^{\dagger}_{i,\sigma} f_{i+\delta, \sigma} \v b^{\dagger}_{i+\delta}\  \mathcal C_6^{-1}  \Upsilon  \mathcal C_6\  \v b_{i}
    \nonumber \\*
    &\qquad\quad +\sum_{\delta = \pm \v e_3} f^{\dagger}_{i,\sigma} f_{i+\delta, \sigma} \v b^{\dagger}_{i+\delta}\  \mathcal C_6^{-2} \Upsilon \mathcal C_6^2\  \v b_{i} \Bigg] 
    \nonumber \\*
    &- \sum_{a \in (\text{V ex.})} (-1)^{P_{a,{\rm v}}} J_{a,{\rm v}} \, \big( \mathcal{\hat P}_{a,{\rm v}} + \mathcal{\hat P}_{a,{\rm v}}^{-1} \big)
    \nonumber \\*
    &+ \ \cdots \ + \ \left [U=\infty\right ],
\end{align}
%
%
%
%
%
where $\v e_1 = [1,0]$, $\v e_2 = [1/2,\sqrt 3 /2]$, $\v e_3 = [-1/2,\sqrt 3 /2]$, and $f_{i,\sigma}^{\dagger}$ is again the creation operator of an electron living at the WC site $i$.
The first term describes on-site orientation-mixing processes corresponding to $\Delta$; the second term describes vacancy hopping processes in the $\pm \v e_{1}$ directions; and the third (fourth) term describes vacancy hopping processes in the $\pm \v e_{2}$ ($\pm \v e_{3}$) directions, which can be related to the second term by $\mathcal C_6$ ($\mathcal C_6^2$) rotation [see Fig.~\ref{fig:Tunneling}(e)]. In the fifth term, $a \in (\rm{V\ ex.})$ denotes one of the exchange processes around a vacancy shown in Fig.~\ref{fig:Tunneling}(d).
The omitted terms correspond to hopping and exchange processes other than those shown in Fig.~\ref{fig:Tunneling}(d-e) and direct (elastic) interactions between vacancies \cite{footnote:interaction}. 
Figures ~\ref{fig:matrix_elements}(c-d) shows the hopping matrix elements ($t$) and exchange coefficients ($J$) for processes involving a vacancy calculated from the semi-classical expression~\eqref{eq:tunnel splitting}.

\subsection{Two-Dimensional Bose Gas}
\label{sec:Two-Dimensional Bose Gas}

For Coulomb-interacting bosonic particles, one merely needs to substitute $(-1)^{P_a} \to +1$ in $H_{\rm eff}^{\rm wc}$ (\ref{eq:WC exchanges}) without changing the forms of $H_{\rm eff}^{\rm i}$ and $H_{\rm eff}^{\rm v}$ [\eqref{eq:interstitial Hamiltonian} and \eqref{eq:vacancy Hamiltonian}].
The consequence is that all ring-exchange processes and interstitial and vacancy hopping processes mediate ferromagnetism.
This is a special case of a more general result that the ground state of an interacting multi-component bosonic system is a fully polarized ferromagnet~\cite{Lieb2002bosonFerromagnet,yang2003bosonFerromagnet}.

\section{A single defect: \\ Exact diagonalization study}
\label{sec:Energy of WC defects and the self-doping instability of a WC}

\begin{figure}[t]
    \centering
     \hspace{-5mm}
\includegraphics[width = 0.48\textwidth]{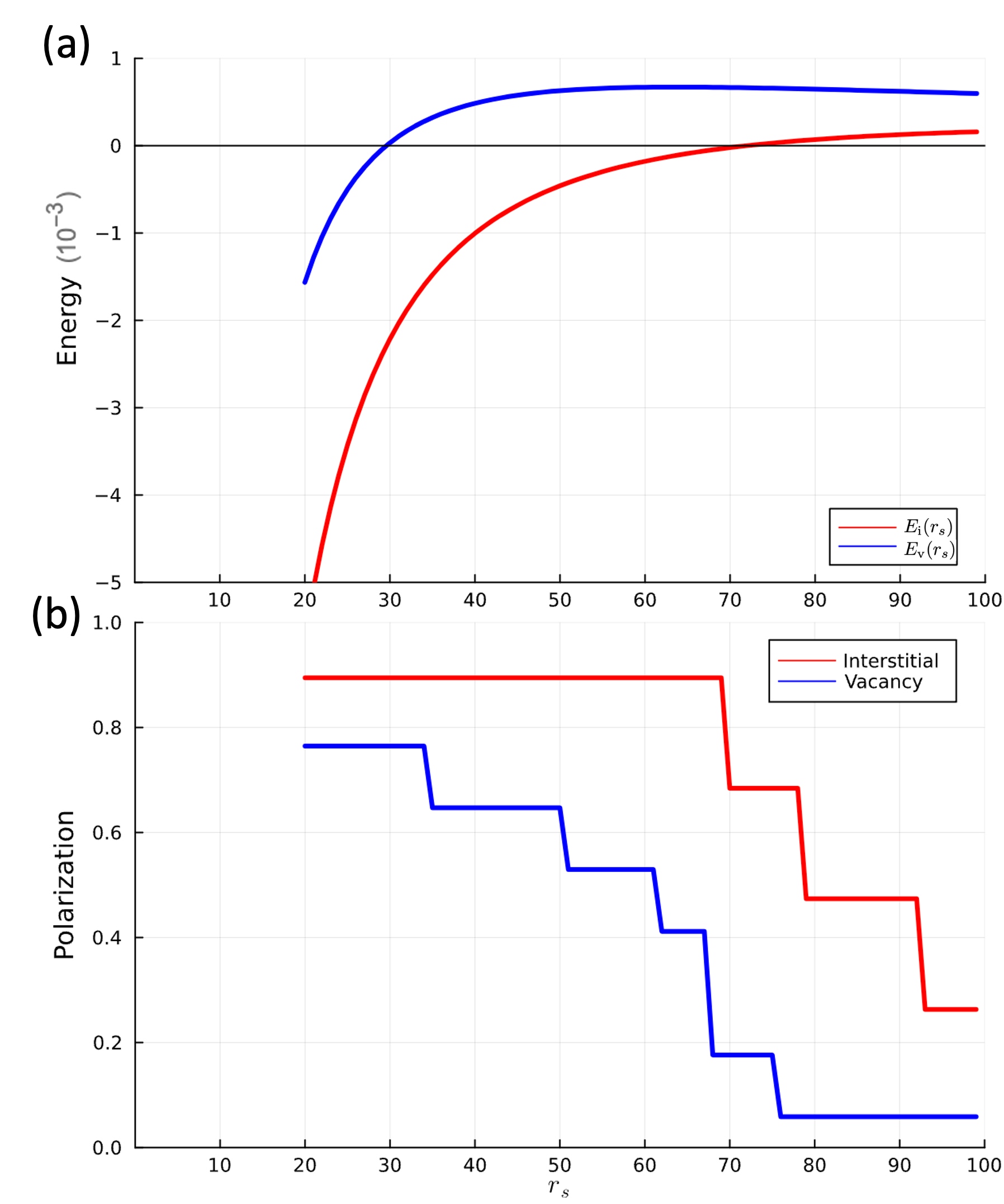}
    \caption{Exact diagonalization results for the effective Hamiltonians (\ref{eq:interstitial Hamiltonian}, \ref{eq:isotropic vacancy Hamiltonian}) on a $3 \times 6$ triangular lattice WC with periodic boundary conditions in the presence of a single defect. 
    The semi-classical expressions (Fig.~\ref{fig:matrix_elements}) are used as an input for the various matrix elements.
    (a) The ground state energy in the presence of a single interstitial ($E_{\rm i}$) and a vacancy ($E_{\rm v}$) as a function of $r_s$ from the resulting semi-classical expressions 
    (\ref{eq:interstitial energy}, \ref{eq:vacancy energy}). 
    $E_{\rm i}(r_s)$ [$E_{\rm v}(r_s)$] crosses zero around $r_s = r_{\rm mit} \approx 70$ [$r_s \approx 30$].
    (b) The relative spin polarization ($0 \leq 2 S^z_{\rm tot}/{N_e} \leq 1$) induced by a single interstitial and vacancy.
    }
    \label{fig:energies}
\end{figure}

In this section, we present the results of a finite-size exact diagonalization study (up to $3\times6 \pm 1$ electrons) of the derived effective Hamiltonians in the single-defect sector.(\ref{eq:interstitial Hamiltonian},\ref{eq:vacancy Hamiltonian})
\footnote{We note that the semi-classical expressions for many dynamical processes considered in this paper may not be taken at their face value in the range $r_s \lesssim 100$, as the instanton approximation breaks down when $\sqrt {r_s} S_a \lesssim 1$.
For example, for $S_{t_1} = 0.098$ and $r_s =100$, $\sqrt{r_s}S_{t_1} = 0.98 \lesssim 1$.
}.
Figure \ref{fig:energies} summarizes the result of the exact diagonalization calculation.

The maximum kinetic energy gain for an interstitial is calculated by obtaining the ground state of $H_{\rm eff}^{\rm i}$ \eqref{eq:interstitial Hamiltonian} in the single-interstitial sector.
We retained all the terms shown in Fig.~\ref{fig:Tunneling}(b-c) except for $J''_{3,{\rm i}}$ and $J_{4,{\rm i}}$. 
(In the range of $20 \leq r_s \leq 100$ considered, they are more than an order of magnitude smaller than the dominant terms in the Hamiltonian.)
The resulting kinetic energy gain, $E^{\rm kin}_{\rm i}(r_s)<0$, is calculated for a system of $3 \times 6$ WC sites with an additional interstitial (i.e., a total of $3\times6+1$ electrons) with periodic boundary conditions.
Including the classical Coulomb energy and the zero-point vibrational energy, the minimum interstitial energy (in units of $e^2/4\pi \epsilon a_{\rm B}$) is 
\begin{align}
\label{eq:interstitial energy}
    E_{\rm i}(r_s) = \frac{C_{1,{\rm i}}}{r_s} 
    + \frac{C_{ 3/2, {\rm i}}}{r_s^{3/2 }}
    + E^{\rm kin}_{\rm i}(r_s) + \cdots.
\end{align}
Here, we have neglected terms corresponding to higher order perturbative corrections (i.e, higher powers of $r_s^{-1/2}$) from phonon anharmonicity as well as higher order corrections to the semi-classical instanton approximation.
We calculated $C_{1,{\rm i}}$ and $C_{3/2,{\rm i}}$ for 
supercells up to size $28 \times 28 + 1$; extrapolation to an infinite supercell size gives $C_{1,{\rm i}} = 0.0769$ and $C_{3/2,{\rm i}} = -0.295$~%
\footnote{
Our results for $C_{1,{\rm i}}$ and $C_{3/2,{\rm i}}$ agree with Ref.~\cite{Cockayne1991defect}.
Note the factor $2$ difference between our values and those of Ref.~\cite{Cockayne1991defect} due to the difference in the unit of energy.}.
The semi-classical expression for the interstitial energy \eqref{eq:interstitial energy} is plotted as a function of $r_s$ in Fig. \ref{fig:energies}(a).

For the vacancy, the on-site orientation-mixing term  $\Delta$ is the largest energy scale in the range of $20 \leq r_s \leq 100$, as shown in Fig.~\ref{fig:matrix_elements}(c-d).
Therefore, we simplify the vacancy problem by projecting it into the ``isotropic single vacancy sector,'' whose basis states are equal superpositions of all the vacancy orientations $\alpha$ at a site $i$:
\begin{align}
    \left | i; \{\sigma \}_{\rm wc}  \right > 
    &\equiv \frac 1 {\sqrt 3} \sum_{\alpha = 1}^3 
    \left | i,\alpha; \{\sigma \}_{\rm wc}  \right > \\*
    &= \frac 1 {\sqrt 3} \sum_{\alpha = 1}^3 
    b^{\dagger}_{i, \alpha} \,
    f^\dagger_{1,\sigma_1} \cdots \bcancel{f}^\dagger_{i,\sigma_i} \cdots f^\dagger_{N,\sigma_N} \vacuum . \nonumber
\end{align}
Here $\{\sigma \}_{\rm wc}$ are the spins of the WC electrons, and the slash in $\bcancel{f}_{i,\sigma_i}$ denotes that the corresponding operator is omitted from the product. 
The projection of $H_{\rm eff}^{\rm v}$ \eqref{eq:vacancy Hamiltonian} to the isotropic single vacancy sector is straightforward, and yields
\begin{align}\label{eq:isotropic vacancy Hamiltonian}
    H_{\rm eff}^{\rm v}\vert_{\Delta}
    &= -2\Delta 
    -t_{\Delta}^{\rm eff} \sum_{\left <i,j \right >} (f^{\dagger}_{i,\sigma}f_{j,\sigma} + {\rm H.c.})
    \\*
    &+ \sum_i (1-n_i) \bigg[
    J_{2}^{\rm eff}\sum_{\left <j,k \right >}
    \mathcal{\hat P}_{j,k} 
     - 
    J_{3}^{\rm eff}\!\!
    \sum_{\left <j,k,l \right >}\!
    \mathcal{\hat P}_{ j,k,l} 
    \nonumber \\*
    &\hspace{6.2em} +
    J_{4}^{\rm eff}\!\!
    \sum_{\left <j,k,l,m \right >}\!\!
    \mathcal{\hat P}_{ j,k,l,m } 
    + {\rm H.c.} \bigg] , \nonumber
\end{align}
where
\begin{align}
    t_{\Delta}^{\rm eff} &\equiv 
    \frac 1 3 [1,1,1] \Upsilon \! \begin{bmatrix} 1 \\ 1 \\ 1 \end{bmatrix}
    = \frac 1 3 (t_{11} + 2t_{22} + 4 t_{12} + 2 t_{23}), 
    \nonumber \\*
    J_2^{\rm eff}\! &\equiv \frac 1 3 (J_{2,{\rm v}}\!+ 2 J'_{2,{\rm v}}\! + \frac 1 2 J''_{2,{\rm v}}), 
    \\*
    J_3^{\rm eff}\! &\equiv \frac 2 3 ( J_{3,{\rm v}}\! + J'_{3,{\rm v}}) , \quad
    J_4^{\rm eff}\! \equiv \frac 1 3 J_{4,{\rm v}} . \nonumber 
\end{align}
In the presence of $N_{\rm v}$ isotropic vacancies, one substitutes $2 \Delta \rightarrow 2 N_{\rm v} \Delta$.
The $(1-n_i)$ factor locates the vacancy, where $n_i = \sum_{\sigma} f^{\dagger}_{i, \sigma} f_{i, \sigma}$, and  $\left <j,k \right >$, $\left <j,k,l \right >$ and $\left <j,k,l,m \right >$ are the 2, 3, and 4 sites neighboring the vacancy location $i$ participating in the corresponding ring-exchange processes ($j,k,l,m$ themselves are also nearest neighbors).
$J''_{2,{\rm v}}, J'_{3,{\rm v}},J_{6,{\rm v}}$ are much smaller than other terms in $H^{\rm v}_{\rm eff}$ and hence will be ignored.

By solving $H_{\rm eff}^{\rm v}\vert_{\Delta}$ \eqref{eq:isotropic vacancy Hamiltonian} in the single vacancy sector, we numerically find the minimum possible vacancy kinetic energy, $E^{\rm kin}_{\rm v}(r_s) < 0$, on $3 \times 6$ WC sites in the presence of a single vacancy ($3\times 6-1$ electrons).
The full semi-classical expression, including the Coulomb and zero-point energy, for the vacancy energy (in units of $e^2/4\pi \epsilon a_{\rm B}$) is then 
\begin{align}
\label{eq:vacancy energy}
     E_{\rm v}(r_s) = \frac{C_{1,{\rm v}}}{r_s} 
     + \frac{C_{ 3/2,{\rm v}}}{r_s^{3/2 }}
     + E^{\rm kin}_{\rm v}(r_s) + \cdots,
\end{align}
where, again, the neglected terms correspond to higher order perturbative and non-perturbative corrections.
We calculated $C_{1,{\rm v}}$ and $C_{3/2,{\rm v}}$ for supercells up to $24 \times 24 -1$; extrapolation to an infinite supercell size gives $C_{1,{\rm v}} = 0.1094$ and $C_{3/2,{\rm v}} = -0.368$.
Whereas our result for $C_{1,{\rm v}}$ agrees with Ref.~\cite{Cockayne1991defect} up to the fourth digit, our value for $C_{3/2,{\rm v}}$ is slightly different from theirs~%
\footnote{Our calculation is performed in a hexagonal supercell as opposed to that of Ref.~\cite{Cockayne1991defect} in a rectangular supercell.}.
The semi-classical expression for the vacancy energy \eqref{eq:vacancy energy} is plotted as a function of $r_s$ in Fig. \ref{fig:energies}(a).

Note that due to the presence of small competing exchange interactions of the underlying WC, a single defect will induce magnetism only in a finite region around it, forming a magnetic polaron. 
Such a competition effectively increases the energy of an interstitial and a vacancy by a small amount as measured relative to the energy of the pure WC.

The semi-classical expressions for the interstitial (vacancy) energy vanishes around  $r_s=r_{\rm mit} \approx 70$ ($r_s \approx 30$), indicating a possible instability of the WC to interstitial self-doping for $r_s < r_{\rm mit}$ (Fig.~\ref{fig:energies}).

\section{Intermediate Phases of the 2DEG}
\label{sec:an intermediate Phase of 2DEG}

Our predicted value of $r_{\rm mit} \approx 70$ is larger than $r_{\rm melt}^* \approx 31$, where $r_{\rm melt}^*$ is the value below which---according to existing variational calculations---the energy of the paramagnetic Fermi liquid becomes smaller than the energy of a WC (with a particular assumed antiferromagnetic order) \cite{drummond2009QMC}.
This suggests that there is a range of densities,
$r_{\rm melt}< r_s < r_{\rm mit}$, for which 
a metallic electron crystal (MeC) phase with more than one electron per crystalline unit cell is stable~%
\footnote{The possibility of a quantum crystal in which the particle density per crystalline unit cell is different from an integer value was originally conceived of in Ref.~\cite{andreev1969quantumcrystals} in the context of solid helium.}.
Here, $r_{\rm melt}$ is the value below which the crystalline order vanishes (see Fig.~\ref{fig:phase diagram}).
(We expect $r_{\rm melt} < r_{\rm MIT}^*\approx 31$ because our metallic WC phase is expected to have a lower energy than the pure WC.)
To the best of our knowledge, the proposed MeC phase with more than one electron per crystalline unit cell has not been studied using the variational quantum Monte Carlo method.
A related, but distinct, MeC phase with less than one electron per crystalline unit cell has been studied in Refs.~\cite{falakshahi2005hybrid} and \cite{drummond2009QMC}; however, these studies are in disagreement with each other.

As discussed in the next section, an interstitial forms a large magnetic polaron in a WC, so the MeC phase occurring near $r_{\rm mit}$ (Fig.~\ref{fig:phase diagram}) is expected to be characterized by an anomalously large quasi-particle (interstitial) effective mass.
Such massive magnetic polarons have a tendency to agglomerate,  leading to phase separation \cite{kivelson2022hubbard, emeryKivelson1990phaseSeparation, Altshuler2002phaseSeparation} and rendering the transition at $r_{\rm mit}$ to be first-order~%
\footnote{This argument ignores a subtlety regarding the formation of mesoscale density-modulated  ``microemulsion'' phases in a (possibly narrow) regime about any putative first-order transition in the presence of  Coulomb interactions \cite{spivak2003phase, spivak2004phases, jamei2005universal, spivak2006transport}.}.
Note that for single component (spinless) electrons, polaron formation is not an issue so the self-doping transition may be continuous~%
\footnote{Note, however, that even if the self-doping transition occurs in the spinful problem, it need not occur in the spinless problem if the spinless (or fully-polarized) MeC phase always has higher energy than both the fully-polarized WC and a fully-polarized FL.}.

Finally, the transition to a fully melted Fermi fluid occurs when the energy of the MeC phase crosses that of the liquid phase.  
Existing variational quantum Monte Carlo estimates of the critical $r_s$ involve comparing the energy of the liquid to that of the insulating WC.  
If, as we have suggested, the MeC has lower energy than the insulating WC in an intermediate $r_s$ regime, it would presumably be stable against quantum melting at somewhat higher densities (smaller $r_s$).  
Thus, this carries with it the likely implication that $r_{\rm melt} < r_{\rm melt}^*\approx 31$.

\section{Kinetic magnetism}
\label{sec:Kinetic magnetism}

Here, we discuss the magnetic correlations induced by defect hopping processes~%
\footnote{The potential importance of defect hopping processes in the magnetism of a quantum crystal was pointed out earlier in Refs.~\cite{thouless1965, castaing1979phase, spivak2000ferromagnetism}. For a comprehensive discussion in the context of solid ${}^3$He, see \cite{montambaux1982vacancies}}.

Distinct interstitial hopping terms induce different magnetic correlations in the underlying WC.
The character of the dominant magnetic correlations induced by each hopping process is determined by the parity of the smallest spin permutation it induces \cite{thouless1965}.
For example, by applying $t_2''$ terms twice on the interstitial, one recovers the same charge configuration but with 3 electrons (spins) permuted. 
This is an even permutation and mediates ferromagnetism  as discussed in Sec.~\ref{sec:WC ring-exchange processes}.
Similarly, the smallest permutation that the $t_2$ terms induce involves $7$ electrons (even permutation) and also mediates ferromagnetism. 
On the other hand, the smallest spin permutation induced by $t_2'$ process involves $4$ electrons (odd permutation) and mediates antiferromagnetism. 
The $t_1$ hopping term does not couple with the underlying WC and hence does not induce magnetism by itself.
Taken together, the various hopping terms, in combination with exchange processes $J_{a,{\rm i}}$, lead to a complicated problem with competing magnetic tendencies.

Interestingly, the interstitial dynamics induces non-trivial spin polarization $2 S^z_{\rm tot}/{N_e}$, as shown in Fig.~\ref{fig:energies}(b), where $S^z_{\rm tot}$ is the total $S^z$ quantum number and $N_e$ is the number of electrons in the system.
For $20 \leq r_s < 70 $, the interstitial seems to always favor a single spin-flip in a fully polarized background (this is also true for smaller systems of $3\times 4 +1$ or $3\times 5 +1$ electrons)~%
\footnote{Such a spin-polaron is also shown to occur in the triangular lattice Hubbard model in the presence of a large Zeeman field \cite{davydova2023spinPolaron, batista2018pairing}.}.

In the presence of small antiferromagnetic WC exchange interactions, a single interstitial can only delocalize in a finite region, forming a large magnetic polaron of size $\sim a_0^2 \sqrt{t/J}$ \cite{spivak2000ferromagnetism, kivelson2022hubbard, kim2022interstitial}, where $t$ and $J$ are characteristic values of interstitial hopping matrix elements and WC exchange coefficients, respectively.
At $r_s \approx 45$, we estimate that a single interstitial induces a magnetic polaron involving $\sim 40$ WC spins.

On the other hand, it is known that the dynamics of a single hole in the $U=\infty$ Hubbard model
on a non-bipartite lattice leads to some form of antiferromagnetism \cite{haerter2005kineticAF, sposetti2014kineticAF, zhu2022doped, kim2023SU2, kim2023SUN}; therefore, assuming that the isotropic vacancy is energetically favored, its hopping processes mediate antiferromagnetic correlations around it. 
In the presence of competing exchange interactions of the underlying WC, a vacancy similarly forms a finite-sized antiferromagnetic polaron.

By controlled doping of a WC in the presence of a smoothly varying weak external periodic potential, one can obtain the defect-doped commensurate WC phase as a stable ground state, as the following reasoning shows.
Consider a weak commensurate potential that has minima $-W<0$ at the triangular lattice WC sites.
When the density is tuned away (but not too far away) from the commensurate value, the defect-doped commensurate WC has an energy per electron 
$\Delta E_{\rm comm}/N \approx -W + E_{\rm def} |\delta| + O(W |\delta|, \delta^2)$
as compared to the pure incommensurate WC, where $\delta$ is the ratio of defect electrons to the total number of electrons,
and $E_{\rm def} = E_{\rm i}$ ($  E_{\rm v}$) is the energy of an interstitial (vacancy) defect in the absence of the  external potential.
Therefore, for a range of doping
$ -W/E_{\rm v}<\delta < W/E_{\rm i}$,
the system will form a defect-doped metallic WC phase that is commensurately locked to the external potential.
Such a phase, in turn, is characterized by defect-induced magnetic correlations with much higher energy scales than the exchange processes of the pure WC.
Therefore, one expects that the magnetic energy scale increases as one moves away from the commensurate filling.
Such a proposal may be experimentally tested in certain Moir\'e systems that support a commensurately locked WC phase \cite{Tang2020hubbardsim,Xu2020correlatedinsulators,Regan2020mottandwigner,moralesduran2021mit,yang2023metalinsulator}.

\section{Effects of weak disorder}
\label{sec:Effects of weak disorder}

Before concluding, we remark on the effect of small quenched disorder on the phase diagram (Fig.~\ref{fig:phase diagram}).
Firstly, even weak disorder is expected to destroy any long-range crystalline order; hence all the electronic crystalline states we have discussed are defined only in an approximate sense as short-range ordered states.
Also, the MeC phase is characterized by the reduced density of mobile electrons and their increased effective mass;
hence, even weak disorder is likely to result in strong localization and destroy the metallic character of the phase.
The resulting disorder-induced intermediate insulating phase is characterized by large magnetic energy scales, associated with the dynamical processes of defects.
This may be an explanation for the recently observed insulating phases with much higher magnetic energy than the exchange scales of the pure WC \cite{hossain2020ferromagnetism, kim2021discovery, falson2022competing}. 
Note that such a proposal predicts an exponential reduction of magnetic energy scales with increasing $r_s$ for $r_s > r_{\rm mit}$~%
\footnote{This explanation for unexpectedly robust insulating magnetism is distinct from, although related to, a previous proposal \cite{kim2022interstitial} based on WC-FL puddle formation.}.

\section*{Acknowledgement}
We thank Boris Spivak for initial insights which led to this investigation and Akshat Pandey for collaboration on a previous work.
We appreciate Veit Elser, Brian Skinner and Shafayat Hossain for interesting comments on the draft. 
K-S.K. acknowledges the hospitality of the Massachusetts Institute of Technology, where this work was completed, and thanks
Aidan Reddy, Seth Musser and Yubo Paul Yang for helpful discussions. 
I.E. acknowledges Eugene Demler, Hongkun Park, Jiho Sung, Pavel Volkov, Jue Wang, and Yubo Yang for helpful discussions on related work.
K-S.K. and SAK were supported in part by the Department of Energy, Office of Basic Energy Sciences, Division of Materials Sciences and Engineering, under contract DE-AC02-76SF00515 at Stanford.
I.E. was supported by AFOSR Grant No. FA9550-21-1-0216 and the University of Wisconsin--Madison.
C.M. was supported in part by the Gordon and Betty Moore Foundation's EPiQS Initiative through GBMF8686, and in part by the National Science Foundation under Grants No.~NSF PHY-1748958 and PHY-2309135.
Parts of the computing for this project were performed on the Sherlock computing cluster at Stanford University.

\appendix

\section{Numerical calculations of $S$ and $A$}
\label{app:Details of the numerical calculation of $S_a$ and $A_a$}

In this section, we review a numerical method for calculating $S_a$ (\ref{eq:action}) and $A_a$ (\ref{eq:fluctuation determinant}), 
closely following Ref.~\cite{voelker2001disorder}.
Although we applied the semi-classical instanton calculation to the 2DEG specifically, the method outlined here applies to any system with a general potential $V(\v r)$ with degenerate minima in the semi-classical limit.
We first calculate the instanton action $S_a$ (\ref{eq:action}) by discretizing a tunneling path:
\begin{align}
\label{eqA:classical action}
    S_{a} &= \int_{{\v r}_0}^{{\v r_0'}} d\v r \, \sqrt{2 \Delta V(\v r ) }
    \\*
    &\approx 
    \sum_{k = 1}^{N_{\rm time}}
    \frac 1 2 |\v r_{k} - \v r_{k-1}| 
    \cdot \big  [\sqrt{2 \Delta V(\v r_k )} + \sqrt{2 \Delta V(\v r_{k-1} )} \big  ] \nonumber
    ,
\end{align}
where we defined $ \Delta  V(\v r ) \equiv  V(\v r ) -V(\v r_0)$ and used the semi-classical equation of motion $ \ddot {\v r} = \nabla V(\v r)$. $\v r_0$ and $\v r_0'$ are initial and final minimum configurations of $V$, respectively,
$\v r_k \equiv \v r (\tau_k)$ is the collective coordinate of particles at time $\tau_k$, where $0\equiv \tau_0<\tau_1 < \tau_2 <\cdots < \tau_M \equiv \tilde \beta$, and $\v r_M \equiv \v r_0'$.
In order to make the distances $|\v r_{k} - \v r_{k-1}|$ approximately equal, each $\v r_{k}$ is taken to be constrained in the hyperplane defined by $(\v r_{k} - \v r_0) \cdot ( \v r_0' - \v r_0) = \frac k {M} |\v r_0' -\v r_0|$. 
Numerical minimization of the discretized action (\ref{eqA:classical action}) is performed with a standard optimization package~\cite{optim}.
We will henceforth denote by $\v r_k$ ($k=0,1,..,M$) the optimized tunneling path for the $a$-instanton process.

The fluctuation determinant $A_a$ captures the Gaussian fluctuations around the semi-classical path
\begin{widetext}
\begin{gather}
    A_a = \frac{F'\big [\v r^{(a)}(\tau)\big  ]}{F\big [\v r_0\big ]} = \left [ \frac{ {\rm det}'\big (-\partial_\tau^2 + V''\big [\v r^{(a)}(\tau)\big ]\big )}{\det\big (-\partial_\tau^2 + V''(\v r_0) \big )} \right ]^{-\frac 1 2},
    \label{eqA:fluctuation det}
    \\
    F\big [\v r(\tau)
    \big  ] \equiv
\int_{\delta \v r (0) =0}^{\delta \v r (\tilde \beta) =0} D \delta \v r (\tau)  \exp\left [-\frac 1 2 \int_0 ^{\tilde \beta} 
    \left ( \delta \dot{\v r}(\tau)^2 + \delta \v r(\tau)^{\rm T}V''\big [ \v r  ^{(a)}(\tau) \big ]\delta \v r(\tau)  \right)
    \right ]
    =
    \left <\v 0 \right | \mathcal{T} \exp(-\int_0^{\tilde \beta}d\tau\, h[ \v r^{(a)} (\tau) ]) \left |\v 0 \right >
   , \label{eqA:F}
   \\ 
   h[\v r (\tau)] \equiv -\frac 1 2  \nabla^2 + \frac 1 2  \delta \v r(\tau)^{\rm T} V''\big [\v r (\tau) \big ] \delta \v r(\tau),
   \label{eqA:deltaH}
\end{gather}
\end{widetext}
where $\mathcal{T} \exp(\cdots)$ denotes the imaginary-time-ordered exponential, $\delta \v r (\tau ) \equiv \v r(\tau) -  \v r^{(a)}(\tau) $ is the fluctuation coordinate, and the primed determinant in the first line is again computed with the zero mode omitted.
$\tilde \beta \to \infty$ is implicitly taken in the end in calculating $A_a.$
As discussed below, the calculation of $A_a$ can be done numerically by first computing ${F\big [\v r^{(a)}(\tau)\big  ]}/ {F\big [\v r_0\big ]}$ that includes the zero mode contribution, and then multiplying  by the square root of the smallest eigenvalue (which is exponentially small in $\tilde \beta$) of the operator $-\partial_\tau^2 + V''\big [\v r^{(a)}(\tau)\big ]$.

$F\big [\v r^{(a)}(\tau) \big  ]$ can be calculated by discretizing the path integral expression  (\ref{eqA:F}).
First, we further define the time slices intermediate to those defined above 
\begin{align}
    0<\tau_{1/2}< \tau_1 < \tau_{1+ 1/2} <\cdots <  \tau_{M -  1/2} <\tau_{M} \equiv \tilde \beta,
\end{align}
where each interval,   $\Delta \tau_k \equiv \tau_{k+ \frac 1 2} - \tau_{k -\frac 1 2} $ ($k=1,\cdots, M-1$), is calculated by inverting the semi-classical equation of motion 
\begin{align}
    \Delta \tau_k &\equiv \int_{\v r_{k - \frac 1 2}}^{\v r_{k + \frac 1 2}} \frac {d\v r }{\sqrt{2 \Delta V[\v r^{(a)}(\tau) ]}}
    \nonumber \\
    & \approx
     \frac {1}{\sqrt{2 \Delta V(\v r_k )}}\cdot \frac 1 2 \left ( |\v r_{k+1} - \v r_k| + |\v r_{k} - \v r_{k-1}| \right) ,
\end{align}
and analogously for the  end intervals, 
$\Delta \tau_{0 } \equiv \tau_{\frac 1 2} \approx \frac {1}{\sqrt{2 \Delta V(\v r_0 )}} \cdot \frac 1 2  |\v r_{1} - \v r_0| $ and $\Delta \tau_{M } \equiv \tau_{M} - \tau_{M - \frac 1 2 } \approx \frac {1}{\sqrt{2 \Delta V(\v r_M )}} \cdot \frac 1 2  |\v r_{M} - \v r_{M-1}|$. 
(Note that the end intervals formally diverge, $\Delta \tau_{0,M } \to \infty$, as $\tilde \beta \to \infty$.)
Then, the propagator at each interval can be approximated by that of the quantum harmonic oscillator (Mehler kernel) of  $h[\v r^{(a)} (\tau)] \approx h_k \equiv h[\v r_k]  $
\begin{widetext}
\begin{gather}
 \left < \delta \v r _{k+\frac 1 2 } \right | 
     e^{- \Delta \tau_{k} h_k  }
     \left | \delta \v r _{k -\frac 1 2} \right >  
= \prod_{n=1}^{2N} \left (\sqrt{B^{(a)}_{n,k}} \exp[ - S^{(a)}_{n,k}] \right ), \label{eqA:Mehler kernel}
\\
S^{(a)}_{n,k} = \frac {A^{(a)}_{n,k}} 2 \left [\left <\v v_{n,k} \middle | \delta \v r_{k-\frac 1 2 } \right >^2 + \left <\v  v_{n,k} \middle | \delta \v r_{k+\frac 1 2 } \right >^2 \right ] 
 - B^{(a)}_{n,k} \left <\v  v_{n,k} \middle | \delta \v r_{k-\frac 1 2 } \right >\left < \v v_{n,k} \middle | \delta \v r_{k+\frac 1 2 } \right >,
\end{gather}
\end{widetext}
 where 
\begin{align}
 & V''(\v r_k) \v  v_{n,k} \equiv (\omega_{n,k})^2 \v v_{n,k},\ \ \ (n=1,\cdots, 2N), \label{eqA:normal modes}
 \\
 &
  A^{(a)}_{n,k} \equiv \frac{\omega_{n,k}}{\tanh(\omega_{n,k} \Delta \tau_k)},\ 
 B^{(a)}_{n,k} \equiv \frac{\omega_{n,k}}{\sinh(\omega_{n,k} \Delta \tau_k)}.
\end{align}
Eq.~\ref{eqA:normal modes} defines normal mode frequencies and eigenmodes at each time slice $k$.
Note that at intermediate times $k \neq 0, M$, $\omega_{n,k}$ is in general complex.
At the end intervals $k=0,M$, one substitutes $\delta \v r_{- \frac 1 2 } \to \delta r_{0} =0$ and $\delta \v r_{M+ \frac 1 2 } \to \delta r_{M} =0$ in the above expressions.
(Note  that as $\tilde \beta \to \infty$, the  propagators at the end intervals approach zero exponentially. 
However, as we will see below, such contributions cancel when calculating $A_a$ as we are calculating the ratio between two $F$s.)

$F\big [\v r^{(a)}(\tau)
    \big  ]$ can finally be computed by integrating over the intermediate fluctuation coordinates $\delta \v r_{k- \frac 1 2 }$
    \begin{widetext}
\begin{align}
     F\big [\v r^{(a)}(\tau)
    \big  ] 
    &=
 \int \left ( \prod_{k=1}^{M} d^{2N} \delta \v r_{k- \frac 1 2 } \right )
    \left <  \v 0 \middle | 
     e^{- \Delta \tau_{M} h_M  }
     \middle | \delta \v r_{M -\frac 1 2} \right >  \left <   \delta \v r_{M- \frac 1 2 } \middle | 
      e^{- \Delta \tau_{M-1} h_{M-1}  }
      \middle | \delta \v r_{M -\frac 3 2} \right > \cdots  
     \left <  \delta \v r_{\frac 1 2} \middle | 
     e^{- \Delta \tau_{0} h_0}
     \middle |  \v 0 \right > 
     \nonumber \\*
     &= 
     \left ( \prod_{k=0}^{M}\prod_{n=1}^{2N} \sqrt{B_{n,k}^{(a)}} \right ) \det(\mathcal M^{(a)})^{-\frac 1 2},
     \\
     \mathcal M^{(a)} &\equiv 
     \sum_{k=1}^{M} e_{k,k} \otimes (  \mathcal A_{k-1}^{(a)}+ \mathcal A_k^{(a)} ) 
     -\sum_{k=1}^{M-1} (e_{k,k+1}+ e_{k+1,k}) \otimes   \mathcal B_k^{(a)} 
     \nonumber \\
     &=
     \begin{bmatrix}
         \mathcal A_0^{(a)}+ \mathcal A_1^{(a)} & -\mathcal B_1^{(a)} & \v 0 & \cdots &\v 0 
         \\
         -\mathcal B_1^{(a)} & \mathcal A_1^{(a)}+ \mathcal A_2^{(a)} &
          -\mathcal B_2^{(a)} & \cdots &\v 0
          \\
          \v 0
          & -\mathcal B_2^{(a)}
          &  \mathcal A_2^{(a)}+ \mathcal A_3^{(a)} & \ddots& \vdots
          \\
          \vdots & \vdots &
          \ddots &   \ddots    & -\mathcal B_{M-1}^{(a)}   
          \\
          \v 0   & \v 0 & \cdots & -\mathcal B_{M-1}^{(a)} 
          &\mathcal A_{M-1}^{(a)}+ \mathcal A_{M}^{(a)}
     \end{bmatrix},
     \\
     \mathcal A_k^{(a)} &\equiv \sum_{n=1}^{2N} A_{n,k}^{(a)} \v v_{n,k} \v v_{n,k}^{\rm T},\ 
     \mathcal B_k^{(a)} \equiv \sum_{n=1}^{2N} B_{n,k}^{(a)} \v v_{n,k} \v v_{n,k}^{\rm T}. \label{eqA:tilde A}
 \end{align}
    \end{widetext}
Here,
$\mathcal M^{(a)}$ is a real symmetric block tridiagonal matrix, 
$e_{i,j}$ is the $M \times M$ matrix with $1$ at the $(i,j)$-th entry with all other entries $0$, 
$\otimes$ is the Kronecker product of two matrices
and $\mathcal A_k^{(a)}$ and $\mathcal B_k^{(a)}$  are $2N \times 2N$ matrices. 
[Note that in the present WC problem, one needs to project out  two zero eigen-modes  $\v v_{n,k}$ (for each $k$) corresponding to uniform translations in the $x$ and $y$ directions; hence $\mathcal A_k^{(a)}$ and  $\mathcal B_k^{(a)}$ become $(2N-2) \times (2N-2)$ matrices.]

In calculating $F[\v r_0]$---which essentially is the propagator of a quantum harmonic oscillator---with the same procedure, one merely substitutes $h_k \to h_0$ in every equation Eq. (\ref{eqA:Mehler kernel}--\ref{eqA:tilde A})
\begin{align}
& F[\v r_0] =
     \left ( \prod_{k=0}^{M}\prod_{n=1}^{2N} \sqrt{B_{n,k}^{(0)}} \right ) \det(\mathcal M^{(0)})^{-\frac 1 2}
\\
 &A^{(0)}_{n,k} \equiv \frac{\omega_{n,0}}{\tanh(\omega_{n,0} \Delta \tau_k)}, \ \,
 B^{(0)}_{n,k} \equiv \frac{\omega_{n,0}}{\sinh(\omega_{n,0} \Delta \tau_k)},
\\
     &
     \mathcal A_k^{(0)} \equiv \sum_{n=1}^{2N} A_{n,k}^{(0)} \v v_{n,0} \v v_{n,0}^{\rm T},\ \,
     \mathcal B_k^{(0)} \equiv \sum_{n=1}^{2N} B_{n,k}^{(0)} \v v_{n,0} \v v_{n,0}^{\rm T},
     \\
     &
     \mathcal M^{(0)}  \equiv 
    \sum_{k=1}^{M} e_{k,k} \otimes (  \mathcal A_{k-1}^{(0)}+ \mathcal A_k^{(0)} ) 
     \nonumber \\
    &\ \ \ \ \ \ \ \ \ \ \ \ \ \ \ \ 
     -\sum_{k=1}^{M-1} (e_{k,k+1}+ e_{k+1,k}) \otimes   \mathcal B_k^{(0)}. 
\end{align}
Therefore, 
\begin{align}
    \frac{F\big [\v r^{(a)}(\tau)\big  ]}{F\big [\v r_0\big ]} 
    = 
    \left ( \prod_{k=1}^{M-1}\prod_{n=1}^{2N} {\frac{ B_{n,k}^{(a)}} { B_{n,k}^{(0)}}} \right )^{\frac 1 2} \left [\frac{\det(\mathcal M^{(a)})}{\det(\mathcal M^{(0)})}\right ] ^{-\frac 1 2}. \label{eqA:fluctuation determinant with zero mode}
\end{align}
Here the product over $k$ runs only from $1$ to $M-1$ because the end interval contributions ($k=0,M$)  of $B_{n,k}^{(a)}$ and $B_{n,k}^{(0)}$ are identical although they formally approach $0$ as $\tilde \beta \to \infty$ [since  $\Delta \tau_{0,M} \to \infty$].
Similarly, one takes $A_{n,0}^{(a)} = A_{n,0}^{(0)} = A_{n,M}^{(a)} =A_{n,M}^{(0)} = \omega_{n,0}$ in calculating  $\det \mathcal M$,  as $\Delta \tau \to \infty$ [since $\tanh(\omega_{n,0}\Delta \tau) \to 1$].

Finally, one needs to divide Eq. \ref{eqA:fluctuation determinant with zero mode} by the (formally diverging) zero mode contribution to $F[\v r^{(a)}(\tau)] ={\rm det}\big (-\partial_\tau^2 + V''\big [\v r^{(a)}(\tau)\big ]\big )^{-1/2}$. 
For this, we numerically find the smallest $\lambda$ such that 
\begin{align}
   \frac 1 {F_{\lambda}[\v r^{(a)}(\tau)]} \equiv \det(-\partial_\tau^2 + V''\big [\v r^{(a)}(\tau)\big ] - \lambda)^{\frac 1 2 } = 0,
\end{align}
where the left hand side is calculated similarly as in  Eqs.~(\ref{eqA:Mehler kernel}--\ref{eqA:tilde A})  with the substitution $V''(\v r_k) \to V''(\v r_k) - \lambda.$
The fluctuation determinant is then obtained as 
\begin{align}
    A_a = \sqrt{\lambda}\cdot  \frac{F\big [\v r^{(a)}(\tau)\big  ]}{F\big [\v r_0\big ]}.
\end{align}


\bibliography{apssamp.bib}

\begin{thebibliography}{64}
\bibitem [{foo()}]{footnote:interaction}%
  \BibitemOpen
  \href@noop {} {}\bibinfo {note} {The interaction between defects is small in
  proportion to a polynomial power of $1/d$, where $d$ is the distance between
  them, and can be neglected if the defect concentration is dilute
  enough.}\BibitemShut {Stop}%
\end{thebibliography}

\end{document}